\documentclass{mem}
\usepackage{natbib}\usepackage{txfonts}\usepackage{balance}
\usepackage{graphicx}
\usepackage[a4paper]{hyperref}
\idline{75}{282}
\begin{document}
\def\teff{$T\rm_{eff }$}
\def\kms{$\mathrm {km s}^{-1}$}

\title{
Simulations of Nuclear Cluster formation
}

   \subtitle{}

\author{
P. \,Miocchi
\and R. \, Capuzzo-Dolcetta
          }

  \offprints{P. Miocchi}

\institute{``Sapienza'' Universit\'a di Roma, 
Dipartimento di Fisica, 
P.le A. Moro, 5, I-00185 Roma, Italy.
\email{miocchi@uniroma1.it}
}

\authorrunning{Miocchi \& Capuzzo-Dolcetta }

\titlerunning{Simulations of Nuclear Clusters formation}

\abstract{
Preliminary results are presented about a fully self-consistent $N$-body
simulation of a sample of four massive globular clusters in close interaction 
within the central region of a galaxy. The $N$-body representation
(with $N=1.5\times 10^6$ particles in total) of both the
clusters and the galaxy allows to include in a natural and self-consistent way
dynamical friction and tidal interactions. 
The results confirm the decay and merging of globulars as a viable scenario
for the formation/accretion of compact nuclear clusters.
Specifically: i) the frictional orbital decay is $\sim 2$ times faster than that
predicted by the generalized Chandrasekhar formula; ii) the progenitor clusters
merge in less than $20$ galactic core-crossing time ($t_b$); iii) the NC
configuration keeps a quasi-stable state at least within $\sim 70 t_b$.
\keywords{Stellar Dynamics --  Methods: numerical --
Galaxies: kinematics and dynamics -- Galaxies: star clusters}
}
\maketitle{}

\section{Introduction}
We present the preliminary results of a fully self-consistent $N$-body
simulation concerning the close interaction of a sample of four massive globular clusters
(GCs) in the central region of a galaxy. Both the clusters and the galaxy are represented
by mutually interacting particles, thus including in a natural and self-consistent way
dynamical friction and tidal interactions. 
This study represents a substantial improvement in the analysis of the
frictional decaying and merging of GCs in galactic nuclear regions, a scenario first tackled by
semi-analitical approaches \citep{tos,cd93} and then pursued
by $N$-body experiments \citep{oh00,cdm08a}. 
Clarifying the role of the above-mentioned dynamical effects
is important also to understand the formation and origin of
Nuclear Clusters (NCs) \citep[e.g.][]{oh00,bekki04}.
\section{Methods and Results}
\begin{figure}%[t!]
\resizebox{\hsize}{!}{\includegraphics{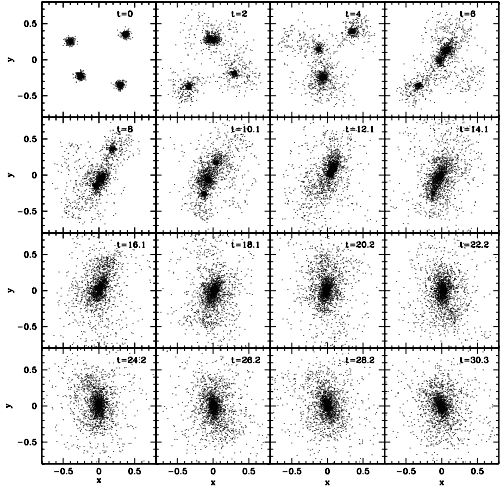}}
%\hfill\includegraphics{miocchi-fig1right.eps}}
\caption{\footnotesize
Time sequence of the merging event of the 4 GCs. Time and lengths are in unit of
$t_b$ and $r_b$.
}
\label{f1}
\end{figure}

Each GC is represented by 256,000 particles initially distributed according to a
King profile whose structural parameters are taken from the set of the
most compact clusters simulated in \citet{cdm08a}.
%total mass $\sim 4.2$--$5.5\times 10^7$ M$_\odot$;
%King and limiting radius $\sim 2$--$4$ pc and $\sim 30$ pc, respectively;
%velocity parameter $\sigma_0 \sim 90$--$100$ km s$^{-1}$;
%central density $\rho_0 \sim 10^5$ M$_\odot$ pc$^{-3}$.
The GCs are initially located at rest within the galactic core (see Fig.~\ref{f1}).
The galaxy is represented by a spherical and isotropic Plummer phase-space distribution
%$f(r,v) \propto (-E)^{7/2}$,
sampled with 512,000 particles.
The $N$-body simulation is performed with our own parallel tree-code using individual and variable
time-steps \citep{mio02}.

The simulation results can be re-scaled with any given set of galactic structural parameters.
One possible choice for these parameters is the following: 
core radius $r_b=200$ pc;
%core mass $M_{bc}=6.7 \times 10^9$ M$_\odot$;
%total mass $=1.9\times 10^{10}$ M$_\odot$;
core-crossing time $t_b=0.54$ Myr;
%central velocity dispersion $\sigma_b=380$ km s$^{-1}$;
central density $\rho_{b0}=370$ M$_\odot$ pc$^{-3}$.

\begin{figure}%[t!]
\includegraphics[width=6.5cm]{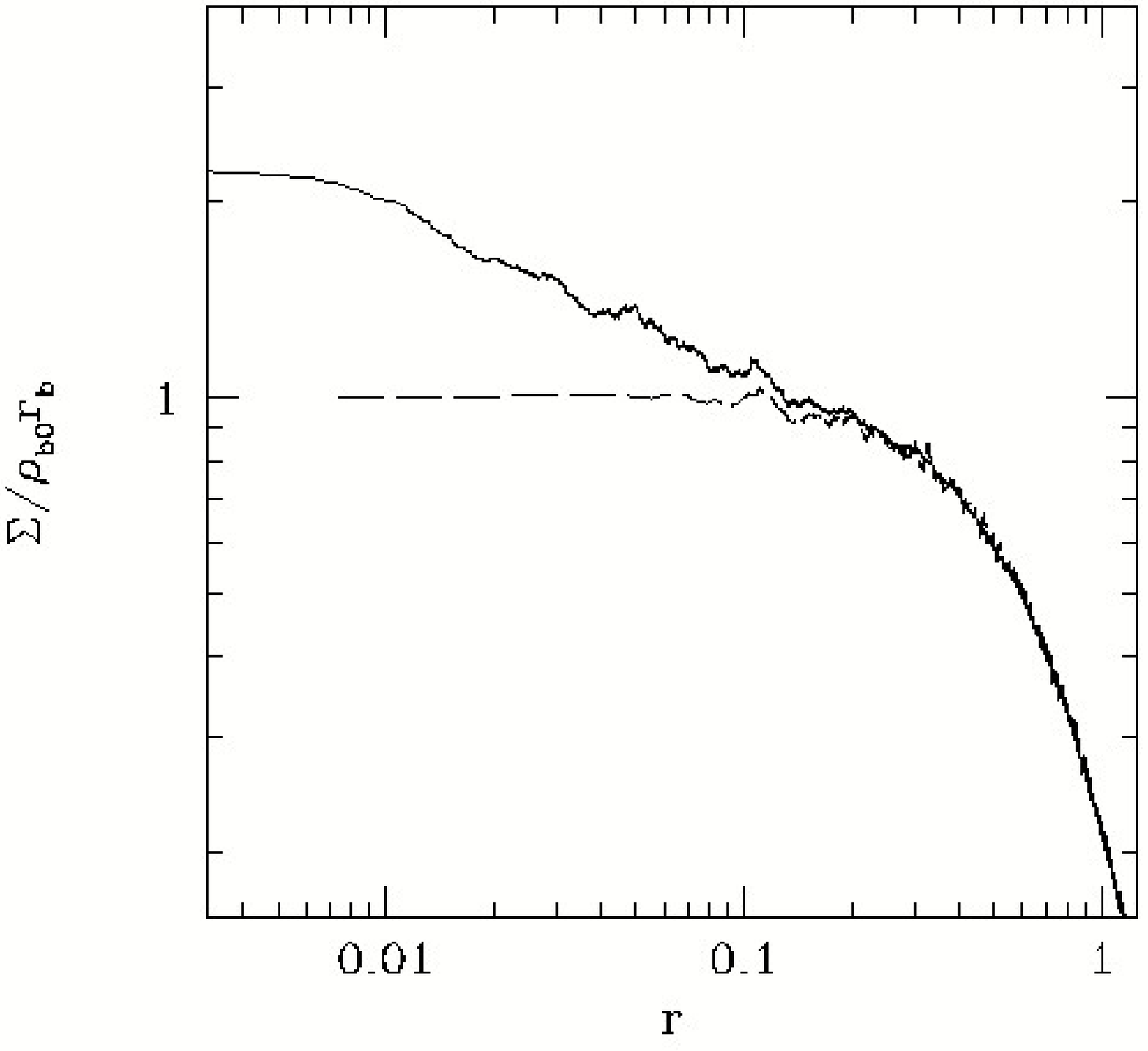}\vfill
\includegraphics[width=6.5cm]{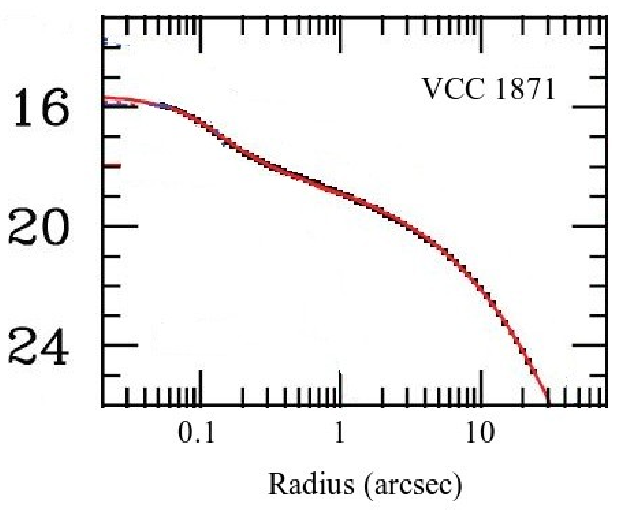}
\caption{\footnotesize
Top: total projected surface density profile (solid line) of  the
last configuration of our NC simulation (the dashed line is the galactic profile only).
$r$ is in unit of $r_b$.
Bottom: the observed profile for the nucleated galaxy VCC 1871 \citep{cote06}.}
\label{f3}
\end{figure}

The main results can be summarized as follows:
i) the frictional orbital decay is $\sim 2$ times faster than that given by the use of
the generalized Chandrasekhar formula \citep{pcv}; ii) the progenitor clusters (initially located within
the galactic core) merge in less than $20$ galactic core-crossing time ($\sim 11$ Myr),
see Fig.~\ref{f1} and \citet{cdm08b}; iii) the NC configuration is quasi-stable at
least within the simulated time ($\sim 70 t_b \sim 40$ Myr);
iv) the total surface density profile has the typical appearance of a nucleated galaxy
central profile (Fig. \ref{f3}); v) the global velocity dispersion profile
\emph{decreases} towards the centre as found in the \citet{geha02} observations.
These results are described in more detail in \citet{cdm08b}.
%
%The main conclusion is that the decay and merging of GCs is a viable scenario
%for the formation/accretion of compact nuclei.

%\begin{figure}[t!]
%\resizebox{\hsize}{!}{\includegraphics{miocchi-fig2.eps}}
%\caption{\footnotesize
%Lagrangian radii (in unit of $R$) time behaviour (top to bottom: 90, 50, 30, 10\%),
%evaluated with respect to the center-of-density of the system made up of the 4 GCs.
%Time is in unit of $t_g$.}
%\label{f2}
%\end{figure}

\begin{acknowledgements}
The simulation was conducted at the CINECA supercomputing centre, under
the INAF-CINECA agreement (grant cne0in07).
\end{acknowledgements}

\bibliographystyle{aa}

\begin{thebibliography}{}
\bibitem[{Bekki et al. (2004)}]{bekki04}Bekki, K., Couch, W.~J., Drinkwater, M.~J.,
Shioya, Y. 2004, \apj, 610, L13
\bibitem[{Capuzzo-Dolcetta (1993)}]{cd93}Capuzzo-Dolcetta, R., 1993, \apj, 415, 616
\bibitem[{Capuzzo-Dolcetta \& Miocchi (2008a)}]{cdm08a}Capuzzo-Dolcetta, R., \& Miocchi, P. 2008a,
\apj, 681, 1136
\bibitem[{Capuzzo-Dolcetta \& Miocchi (2008b)}]{cdm08b}Capuzzo-Dolcetta, R., \& Miocchi, P. 2008b,
\mnras, accepted (astro-ph/0804.4421)
\bibitem[{C\^ot\'e et al. (2006)}]{cote06}C\^ot\'e, P. et al., 2006, \apss, 165, 57
%\bibitem[{Fellhauer et al. (2002)}]{fell02}
%Fellhauer, M., Baumgardt, H., Kroupa, P., Spurzem, R. 2002, Cel. Mech. Dyn. Astron., 82, 113
\bibitem[{Geha et al. (2002)}]{geha02}Geha, M., Guhathakurta, P., \& van der Marel, R.P. 2002, \aj, 124, 3073
\bibitem[{Miocchi \& Capuzzo-Dolcetta, 2002}]{mio02}Miocchi, P., \& Capuzzo-Dolcetta, R. 2002, \aap, 382, 758
\bibitem[{Oh \& Lin (2000)}]{oh00}Oh, K.S., \& Lin, D.N.C. 2000, \apj, 543, 620
\bibitem[{Pesce et al. (1992)}]{pcv}Pesce, E., Capuzzo-Dolcetta, R., \& Vietri, M. 1992, \mnras, 254, 466
\bibitem[{Tremaine et al. (1975)}]{tos}Tremaine, S., Ostriker J.~P., \& Spitzer, L.~Jr. 1975, \apj, 196, 407
\end{thebibliography}

\end{document}